\documentclass[aps,preprint,showpacs,superscriptaddress,groupedaddress]{revtex4}  
\usepackage[dvips]{graphicx}
\usepackage{dcolumn}   
\usepackage{bm}        
\usepackage{amssymb}   
\usepackage[utf8]{inputenc}
\usepackage{mathrsfs}
\usepackage{amsmath}
\usepackage{hyperref}
\usepackage{amsmath}

\usepackage{color}

\begin{document}

\widetext

\title{Search for long-living topological solutions of nonlinear $\varphi^4$ field theory}
\author{Alexander E. Kudryavtsev}
\email{kudryavt@itep.ru}
\affiliation{National Research Center Kurchatov Institute, Institute for Theoretical and Experimental Physics, Bolshaya Cheremushkinskaya street 25, 117218 Moscow, Russia}
\author{Mariya A. Lizunova}
\email{m.a.lizunova@uu.nl}
\affiliation{Institute for Theoretical Physics, Centre for Extreme Matter and Emergent Phenomena, Utrecht University, Princetonplein 5, 3584 CC Utrecht, The Netherlands}
\affiliation{Institute for Theoretical Physics, University of Amsterdam, Science Park 904, 1098 XH Amsterdam, The Netherlands}

\vskip 1cm

\begin{abstract}
We look for long-living topological solutions of classical nonlinear $(1+1)-$dimensional $\varphi^4$ field theory. To that effect we use the well-known cut-and-match method. In this framework, new long-living states are obtained in both topological sectors. In particular, in one case a highly excited state of a kink is found. We discover several ways of energy reset. In addition to the expected emission of wave packets (with small amplitude), for some selected initial conditions the production of kink-antikink pairs results in a large energy reset. Also, the topological number of a kink in the central region changes in the contrast of conserving full topological number. At lower excitation energies there is a long-living excited vibrational state of the kink; this phenomenon is the final stage of all considered initial states. Over time this excited state of the kink changes to a well-known linearized solution --- a discrete kinks excitation mode. This method yields a qualitatively new way to describe the large-amplitude bion, which was detected earlier in the kink-scattering processes in the nontopological sector.

\end{abstract}



\maketitle


\newpage
\section{\label{sec:sec1} Introduction}
\begin{figure}[h]
\begin{minipage}[h]{0.49\linewidth}
\center{\includegraphics[width=0.83\linewidth]{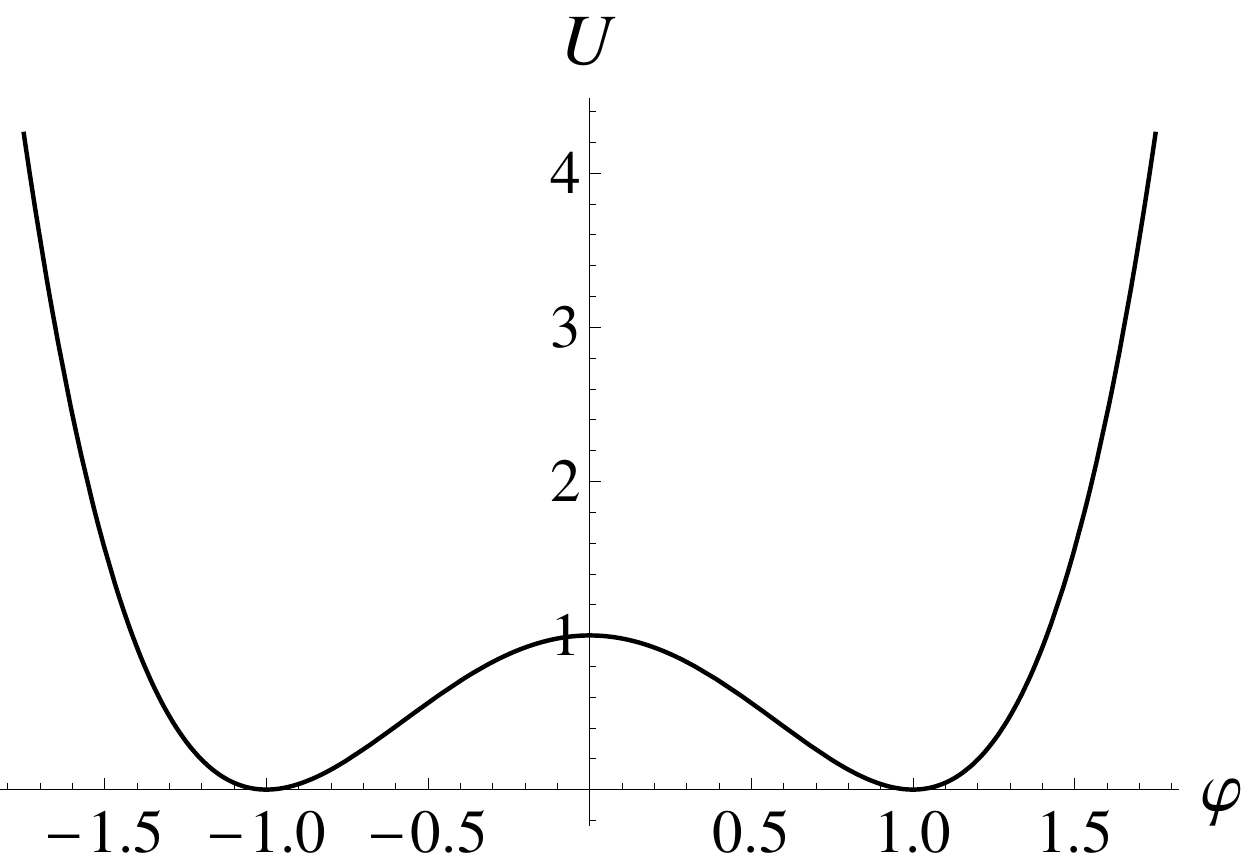}}
\end{minipage}
\hfill
\begin{minipage}[h]{0.49\linewidth}
\center{\includegraphics[width=0.83\linewidth]{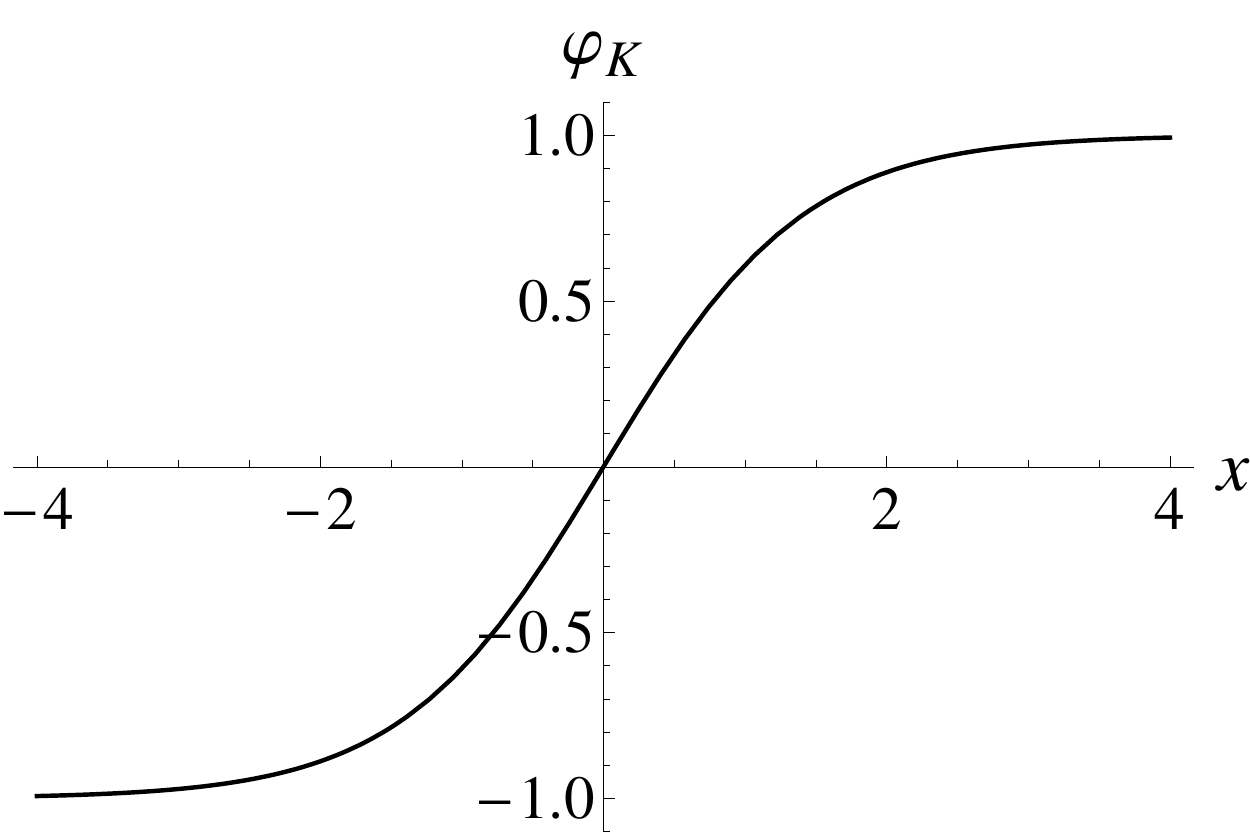}}
\end{minipage}
\caption{The dimensionless potential \eqref{eq:potfi4} of the $\lambda\varphi^4$ model (left panel) and the kink solution [Eq.~\eqref{eq:kinkphi4}] (right panel).}
\label{fig:sec2pic1}
\end{figure}

We consider the $\lambda\varphi^4$ theory with a real scalar field $\varphi(t,x)$ in $(1+1)$ dimensions  \cite{aek,bazeia01,radjaraman}. Its dynamics determined by the following Lagrangian:
\begin{equation}\label{eq:lagrang}
\mathscr{L}=\frac{1}{2} \left( \frac{\partial\varphi}{\partial t} \right)^2-\frac{1}{2} \left( \frac{\partial\varphi}{\partial x} \right) ^2-U(\varphi),
\end{equation}
where $U(\varphi)$ is a potential, defining the self-interaction of the field in the considered model \cite{aek},
\begin{equation}\label{eq:potfi4}
 U(\varphi)=\frac{\lambda}{4}\left(\frac{m^2}{\lambda}-\varphi^2\right)^2.
\end{equation}
The plot of Eq.~\eqref{eq:potfi4} is shown at Fig.~\ref{fig:sec2pic1} (left panel). We analyze a model with a non-negative potential with two minima, so all static solutions with finite energy split into disjoint classes, so-called topological sectors, according to their asymptotic behavior at very large $x$. Solutions with $\varphi(-\infty)\neq\varphi(+\infty)$ are called topological, while those with $\varphi(-\infty)=\varphi(+\infty)$  are nontopological. Both types of the solutions are of growing interest in physics. In particular, they arise in the questions of three- or two-dimensional domain walls. However, the one-dimensional case also is curious and was considered in different works for diverse models \cite{gani1,gani2,gani3}. In the $\lambda\varphi^4$ model there is a soliton solution called a kink; the phenomenon of ``wobbling kink'' was studied in \cite{barashenkov}, \cite{barlit8}. Moreover, a three- or two-dimensional domain wall presents a one-dimensional kink interpolating two different vacua of the model. In some cases these can be solved approximately \cite{GaKuLi}. The domain walls in the $\lambda\varphi^4$ model can be applied to some cosmological models, for example, during discussions of the dark matter and dark energy \cite{lit4}. The results of numerical simulations in other models \cite{GaKuLi} can be applied to solid-state physics \cite{lit37}. 

The Lagrangian \eqref{eq:lagrang} with \eqref{eq:potfi4} yields the equation of motion for $\varphi(t,x)$. After transition to dimensionless variables it reads
\begin{equation}\label{eq:eom1}
\varphi_{tt}-\varphi_{xx}-\varphi+\varphi^3=0.
\end{equation}
As a next step, we find and study the analytical solutions of Eq.~\eqref{eq:eom1}. Note that the vacua of this model $\varphi_{\scriptsize \mbox{vac}}^{(1)}=-1\mbox{ and }\varphi_{\scriptsize \mbox{vac}}^{(2)}=+1$ are stable solutions of \eqref{eq:eom1}. Moreover, there is the unstable permanent solution $\varphi=0$ with infinite energy.

In addition to the previous solutions, there is also a static, nontrivial, topological, solitary wave-like solution \cite{aek}. It can be easily found by solving the static limit of Eq.~\eqref{eq:eom1},
\begin{equation}	\mbox{K}\equiv\displaystyle\varphi_{\scriptsize \mbox{K}}(x-x_0)=\tanh\frac{(x-x_0)}{\sqrt{2}}.
    \label{eq:kinkphi4}
\end{equation}
The antikink $\overline{\mbox{K}}$ is given by minus $\mbox{K}$. The energy functional for the Lagrangian \eqref{eq:lagrang}, in static case \eqref{eq:kinkphi4}, is called the mass of the kink $M_{\scriptsize\mbox{K}}=2\sqrt{2}/3$. The plot of Eq.~\eqref{eq:kinkphi4} is presented by Fig.~\ref{fig:sec2pic1} (right panel).

\subsection{\label{sec:sec2lev3} Excitation spectrum of the kink}
In order to analyze the excitation spectrum of the static kink, we add to it a small perturbation $\delta \varphi$ to it. In other words, we make the ansatz
\[
\varphi(t,x)=\varphi_{\scriptsize \mbox{K}}(x)+\delta\varphi(t,x)=\varphi_{\scriptsize \mbox{K}}(x)+e^{i\omega t}\psi(x).
\]
By taking the terms in Eq.~\eqref{eq:eom1} linear in $\delta \varphi$, we obtain the following equation:
\begin{equation}\label{eq:hpsiepsi}
\begin{gathered}
\hat{H}\psi=E\psi,\quad \hat{H}=-\frac{d^2}{dx^2}-3\cosh^{-2}\frac{x}{\sqrt{2}}, \\
\quad E=\omega^2-2.
\end{gathered}
\end{equation}
The eigenvalue $\omega_0=0$ belongs to the discrete part of the excitation
spectrum \eqref{eq:hpsiepsi} \cite{aek}, but also there is one vibrational excitation given by
\begin{equation}\label{eq:kinkpsi1}
\begin{gathered}
\delta\varphi_1=\psi_1(x)e^{i\omega_1 t},\quad \omega_1=\sqrt{3/2},\\
\psi_1(x)=\left(\frac{3}{2\sqrt{2}}\right)^{1/2}\tanh\frac{x}{\sqrt{2}}\cosh^{-1}\frac{x}{\sqrt{2}}.
\end{gathered}
\end{equation}

\subsection{\label{sec:sec2lev4} Analytical solution, depending on $\boldsymbol{x}$}
 
The above solutions are not a full set of solutions to the $\varphi^4$ model. Let us consider a static wave solution with infinite energy. We consider the static limit of Eq.~\eqref{eq:eom1}
\begin{equation}\label{eq:eom_static}
\varphi_{xx}=-\varphi+\varphi^3.
\end{equation}
\begin{figure}
\begin{center}
	\includegraphics[scale=0.5]{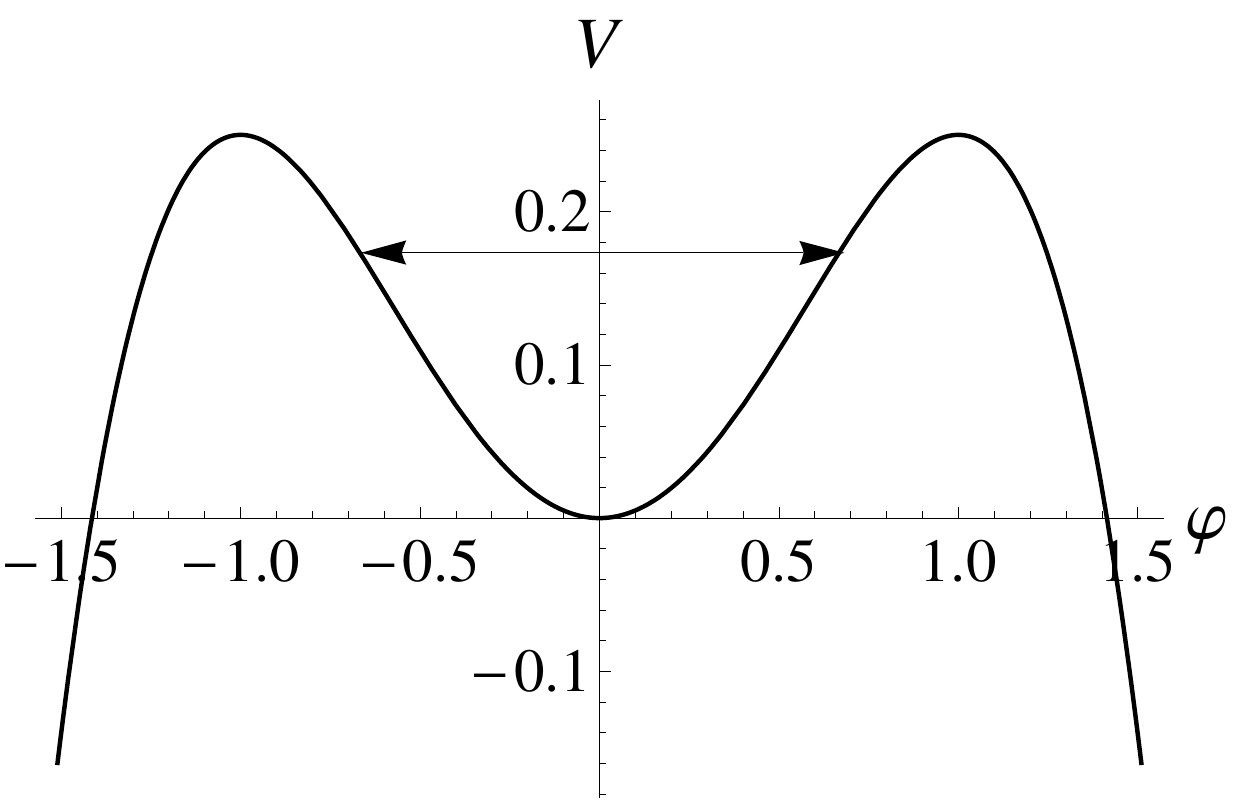}
\end{center}
	\caption{The potential in Eq.~\eqref{eq:v_xvarphi}. The arrow shows the domain of the solution $\varphi(x)$ for an arbitrarily chosen value $0<\varphi_0<1$.}
\label{fig:sec2lev4pic1}
\end{figure}
This equation is analogous to Newton's equation
\begin{equation}\label{eq:v_xvarphi}
\ddot{x}=F(x)=-\nabla V(x),\quad \mbox{where}\quad V(x\to\varphi)=\frac{\varphi^2}{2}-\frac{\varphi^4}{4}+const.
\end{equation}
Figure \ref{fig:sec2lev4pic1} shows a plot of a $V(x\to\varphi)$. In this case $\varphi(x)$ describes a trajectory of an oscillatory movement between points $-\varphi_0$ and $\varphi_0$. In the range $0 < \varphi_0 < 1$ oscillations are periodic. In the limiting case $\varphi_0=1$ the fluctuations disappear, because the time necessary to return to the starting point $\varphi_0=1$ reaches infinity.

\begin{figure}[h]
\begin{minipage}[h]{0.49\linewidth}
\center{\includegraphics[width=0.83\linewidth]{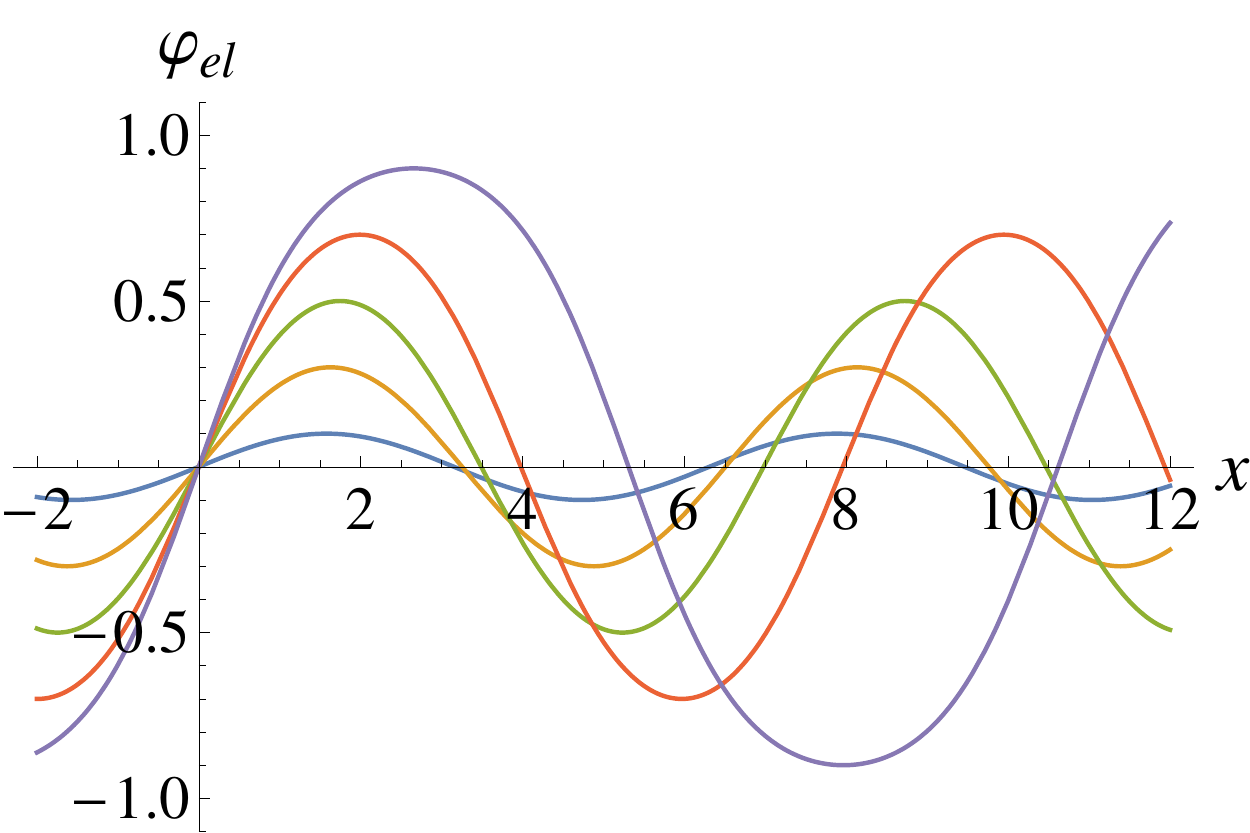}}
\end{minipage}
\hfill
\begin{minipage}[h]{0.49\linewidth}
\center{\includegraphics[width=0.83\linewidth]{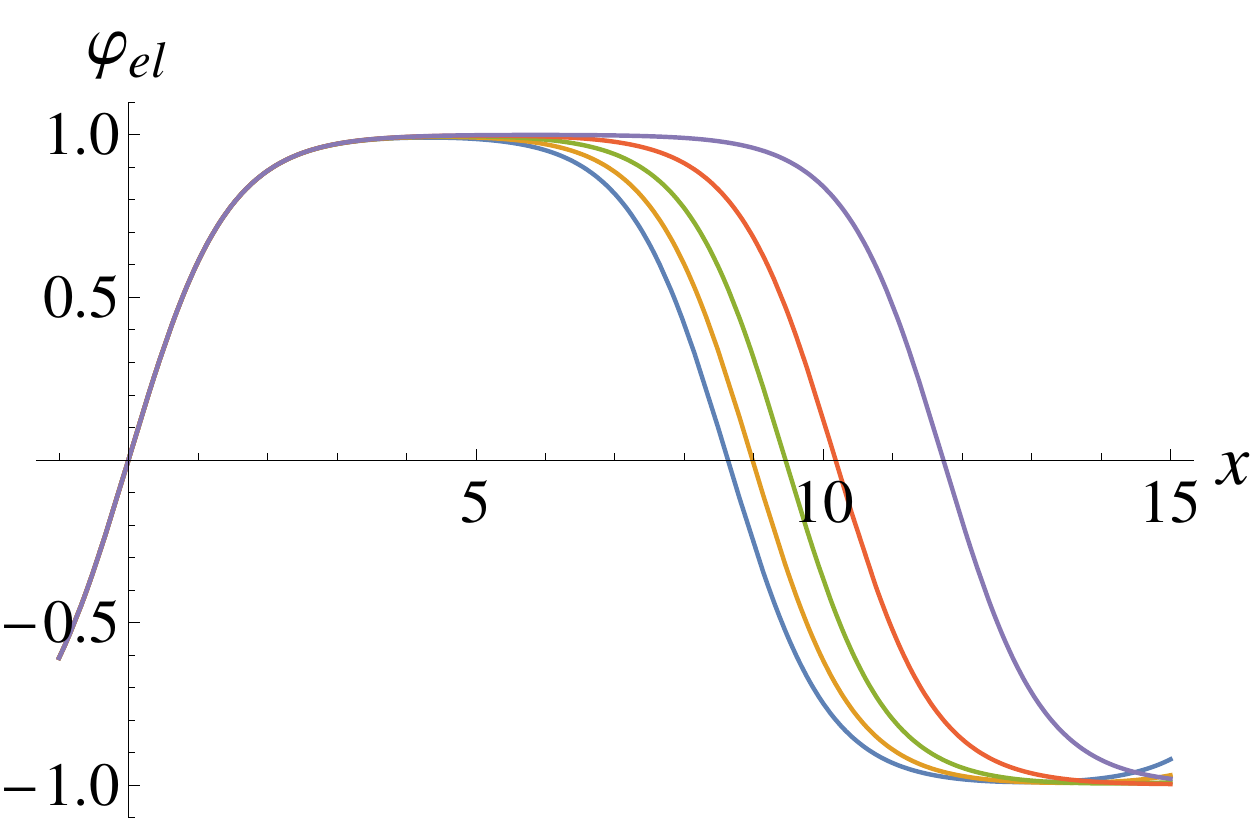}}
\end{minipage}
\caption{The dependence of the solution $\varphi_{\scriptsize \mbox{el}}(x)$ on the parameter $\varphi_0$. The plots are for different $\varphi_0$ from $0.1$ to $0.9$ with step $0.2$ (left panel) and $\varphi_0$ from $0.991$ to $0.999$ with step $0.002$ (right panel). The curves on the plots are ordered by the degree to which the parameter $\varphi_0$ grows.}
\label{fig:sec2lev4pic2}
\end{figure}

Defining the dimensionless variable $\varphi(x)=\varphi_0 \chi(x)$ and the constants
\[
k^2=\frac{\varphi_0^2/2}{1-\varphi_0^2/2},\quad b^2=1-\frac{\varphi_0^2}{2},
\]
where $0\le k^2\le 1$ and $1/2\le b^2\le 1$, leads us to
\[
\begin{gathered}
\int_0^{\chi(x)}\frac{d\chi}{\sqrt{(1-\chi^2)\left(1-k^2\chi^2\right)}}\\
=\langle\chi =\sin\psi \rangle = \int_0^{\arcsin \chi}\frac{ d\psi}{\sqrt{\left(1-k^2\cos^2\psi\right)}}.
\end{gathered}
\]
The last integral is nothing but the elliptic integral of the first kind [$\mbox{F}(\arcsin\chi,k) = b x$] \cite{yanke}. Then, the static periodic solution of Eq.~\eqref{eq:eom_static} can be written as
\begin{equation}\label{eq:varphiel}
\varphi_{\scriptsize \mbox{el}}(x)=\varphi_0~\mbox{sn}(bx,k),
\end{equation}
where $\mbox{sn}(bx,k)$ is the elliptic sine \cite{yanke}. At small $k$ (corresponding to $\varphi_0\ll 1$) there is a concordance $\mbox{sn} (z) \approx \sin (z)$. At $\varphi_0\to 0$ the solution \eqref{eq:varphiel} becomes a permanent unstable solution $\varphi\sim 0$, as previously noted.  Plots of Eq.~\eqref{eq:varphiel} are shown in Fig.~\ref{fig:sec2lev4pic2} for different values of the parameter $\varphi_0$. The elliptic sine period is calculated using the following formula \cite{yanke}:
\begin{equation}\label{eq:varphielT}
T=\frac{4\mbox{F}(\pi/2,k)}{\sqrt{1-0.5\varphi_0^2}}.
\end{equation}

\section{\label{sec:sec3} Formulation of the problem}

The long-living solutions with high amplitude are of growing interest in classical field theory. This type of solution, called bion or breather, was found early in the kink-antikink collisions in the $\varphi^4$ model both in one- and three- dimensional cases \cite{lit38,barlit5f,aeklit65,Manton1,aeklit64,okyn}.

\begin{figure}
\begin{center}
	\includegraphics[scale=0.5]{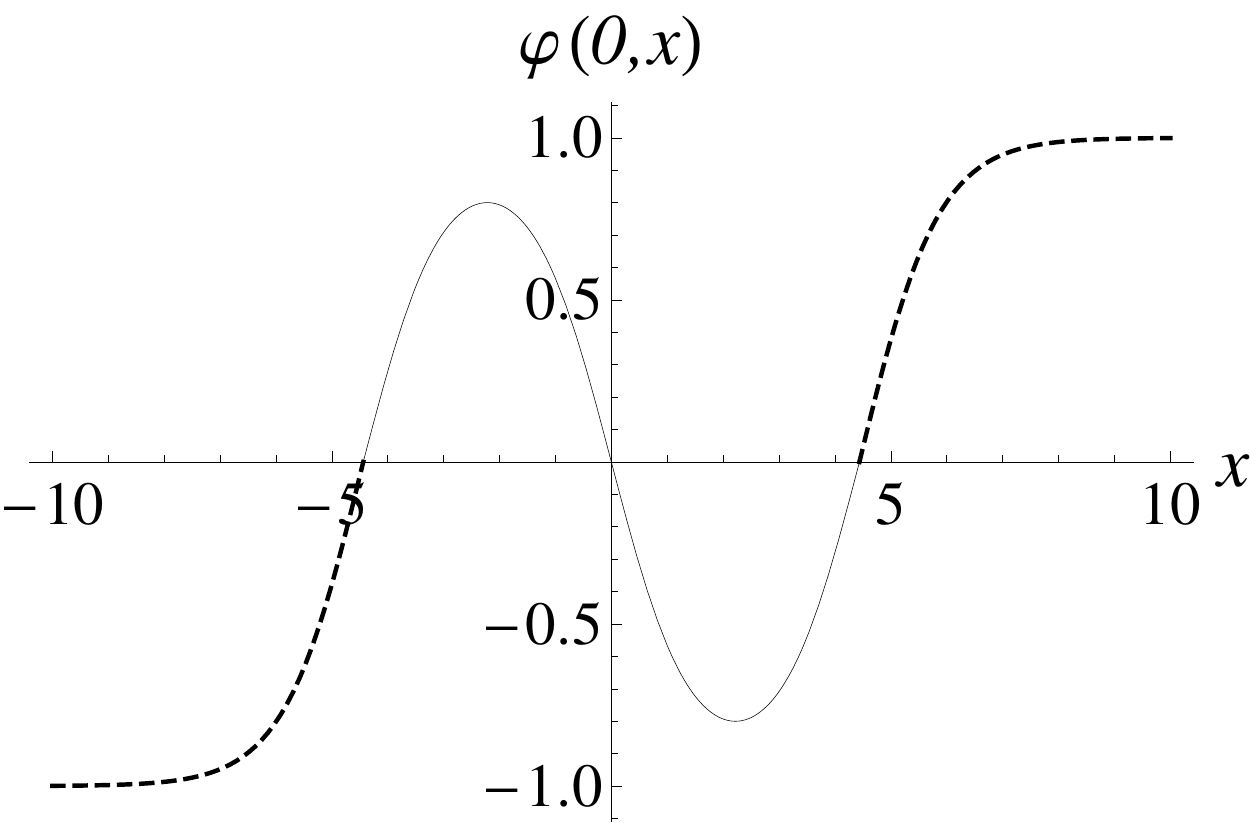}
\end{center}
	\caption{The plot of the initial state, which is constructed with the method ``cut and match''. A dashed line shows the half-kinks \eqref{eq:kinkphi4}, a solid line shows a solution in terms of elliptic function $\varphi_{\scriptsize \mbox{el}}$ for $\varphi_0=0.8$.}
\label{fig:sec3pic1}
\end{figure}

Here, we propose to use the popular cut-and-match method to find a long-living field configuration, using previously found solutions $\varphi=0$, Eq.~\eqref{eq:kinkphi4}, and Eq.~\eqref{eq:varphiel}. In this case, a part of the initial state is composed by the kink \eqref{eq:kinkphi4}, which is divided in two equal pieces at $x=0$. These halves of the kink are fixed at $\pm x_0$. Then, one of the solutions ($\varphi=0$ or $\varphi_{\scriptsize \mbox{el}}$ on the finite interval) of \eqref{eq:eom1} is placed in the space between these two halves. An initial state constructed in manner described is shown in Fig.~\ref{fig:sec3pic1}.

Note that if we take $\varphi=0$, the initial state will become unstable. Its energy linearly increases with growing distance $2x_0$.

In another case we make a solution in terms of the elliptic solution \eqref{eq:varphiel} for a fixed value $0<\varphi_0<1$. For a smooth gluing of selected solutions one defines the value of $x_0$ as a half of the period $T$ of the elliptic function $\varphi_{\scriptsize \mbox{el}}$. Thus we obtain an initial configuration
\[
(-1,\varphi_0,0,-\varphi_0,1),
\]
which we define to mean the following: in the area $-\infty<x<-T/2$ the initial state consists of the left half of \eqref{eq:kinkphi4}, in $-T/2<x<+T/2$ it is given \eqref{eq:varphiel} [such that $\varphi(x=-T/4)=\varphi_0$, $\varphi(x=0)=0$, and $\varphi(x=T/4)=-\varphi_0$], and in the area $T/2<x<\infty$ the solution consists of the right part of \eqref{eq:kinkphi4}. There $T$ is the period of the elliptic function \eqref{eq:varphielT}. The profile of this type of initial state is shown in Fig.~\ref{fig:sec3pic1} (for $\varphi_0=0.8$). First, we consider a ``static initial state'' ($\partial_t\varphi_0=0$), but later we take into account some configurations with dynamics defined by
\[
\frac{\partial\varphi_0(0,x)}{\partial t}=\frac{\varphi_0(\tau,x)-\varphi_0(0,x)}{\tau}=\delta\varphi_0\neq 0.
\]

\subsection{\label{sec:sec3lev1} Numerical solution of the equation of motion}

We solve the partial differential equation \eqref{eq:eom1} using a convergent difference scheme and with nonfixed boundary conditions, while derivatives are approximated by finite differences. The steps are taken as $h=0.04$ (space step) and $\tau=0.02$ (time step), while the equation is solved from $t=0$ to $t=100$. This choice of steps helps to optimize a ratio accuracy of the obtained results and the duration of computing. During the evolution a check of the conservation of energy is performed by taking into account a flow of energy from fixed boundaries. The initial states are compiled with the use of the computer algebra system {\it Mathematica} 8. 

\subsection{\label{sec:sec3lev1} Result for unstable vacuum $\boldsymbol{\varphi=0}$}

The initial condition consists of two halves of the kink, placed in $\pm x_0$, and the unstable zero solution $\varphi=0$ between them. The energy increases linearly with growing value of $x_0$. Two parts in the evolution are observed. First, there is a convergence of both halves of kink with velocity equal to the speed of light. When the halves finally meet each other, two processes alternate: a formation of loops and an emission of waves from the kink (the so-called, ``wobbling kink''). The obtained solution $\varphi_{\scriptsize \mbox{sol}}$ is close to the linearized solution of Eq.~\eqref{eq:eom1}, where $\varphi_{\scriptsize \mbox{sol}}\approx\varphi_{\scriptsize \mbox{K}}+\delta\varphi$, which is very long lived and is characterized by small emission of waves. These waves carry off some energy from the area of localization. Let us explain this phenomenon. At small values of $k(\varphi_0)$ the solution changes from $\varphi_{\scriptsize \mbox{el}}\approx \mbox{sn} x$ to $\varphi\approx \sin x$. As the $\sin$ is a periodic function, when $2x_0\le 2\pi$ the initial condition (with loops) does not cause an excitation like an excited mode of an elliptic function, but instead an excitation like a high-amplitude vibration of a kink. So, the evolution of the initial state can be described qualitatively by
\begin{equation}
\varphi_{\scriptsize \mbox{sol}}\approx \displaystyle\tanh\left(\frac{x}{\sqrt{2}}\right)\left(1+\frac{A(t)}{\cosh\left(x/\sqrt{2}\right)}\right),\quad A(t)=A_0\cos \omega t.
\label{eq:phiwithA}
\end{equation}
The evolution can be described by this equation as there are two modes in the kink spectrum. One of them, which correlates with Eq.~\eqref{eq:kinkpsi1}, is responsible for small vibrations across the solution. We have an idea, that even if the observed vibrations stop being small, they still can be described with a periodic function like the $\cos \omega t$ [we take $\omega=\sqrt{3/2}$ like in Eq.~\eqref{eq:kinkpsi1}]. In Eq.~\eqref{eq:phiwithA} the parameter $A_0$ is taken constant, but it is not a constant in the numerical simulations because there is a small emission from area of localization of the solution.

Moreover, there is one precondition to describe qualitatively an obtained solution accurately by Eq.~\eqref{eq:phiwithA}. The function \eqref{eq:phiwithA} equals zero in $x=0$ one time if $A_0\ge -1$, and three times if $A_0<-1$. Note that a quasiperiodic formation of the loops with period equals $\approx 2\pi$; it is also one reason for using the proposed phenomenological description. This period correlates with $\cos\omega t$,
\[
T=\frac{2\pi}{\omega}\approx 2\pi,\quad\mbox{as}~\omega=\sqrt{\frac{3}{2}}\approx 1.
\]

\begin{figure}[h]
\begin{minipage}[h]{0.49\linewidth}
\center{\includegraphics[width=0.83\linewidth]{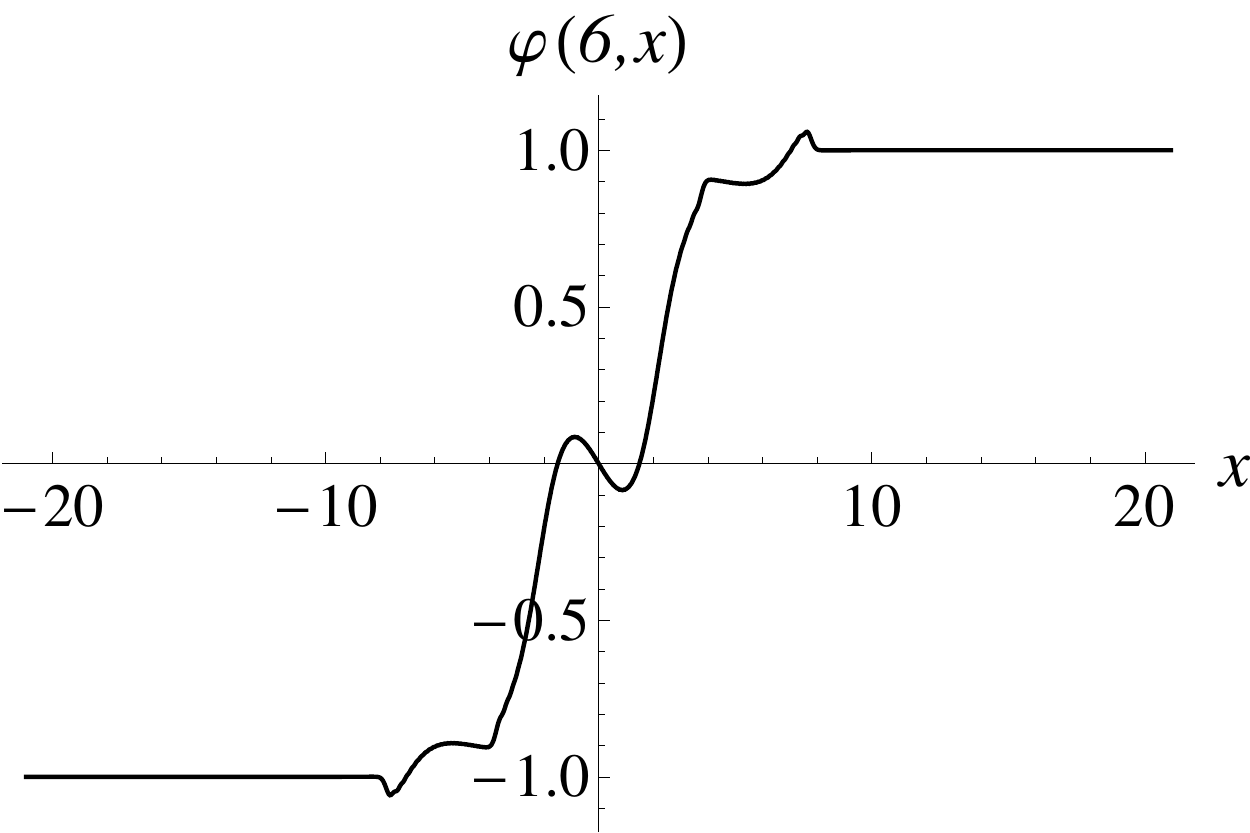}}
\end{minipage}
\hfill
\begin{minipage}[h]{0.49\linewidth}
\center{\includegraphics[width=0.83\linewidth]{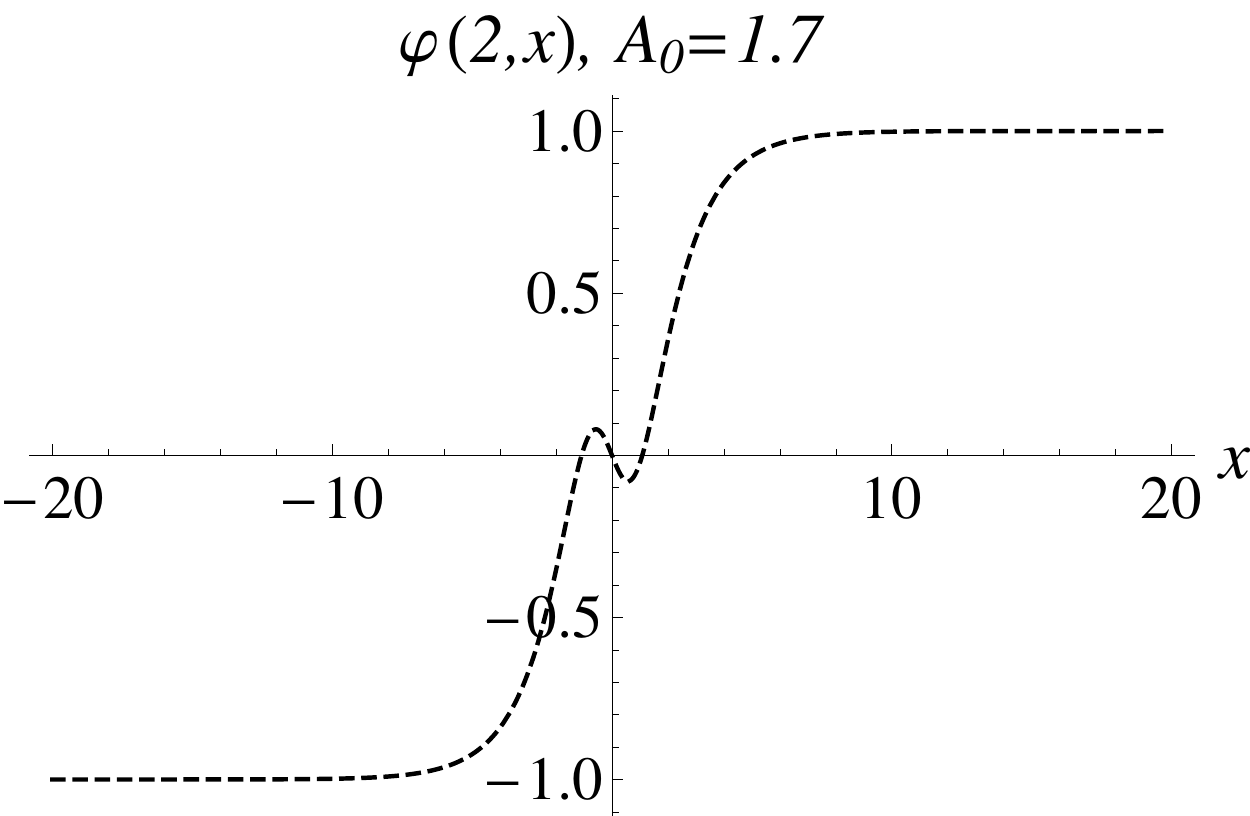}}
\end{minipage}
\caption{The profile of $\varphi(t,x)$ at $t=6$ (left panel) and mapped plot of the solution \eqref{eq:phiwithA} for $A_0=1.7$ at $t=2$ (right panel), $x_0=2$.}
\label{fig:sec3lev1pic3}
\end{figure}

\begin{figure}[h]
\begin{minipage}[h]{0.49\linewidth}
\center{\includegraphics[width=0.83\linewidth]{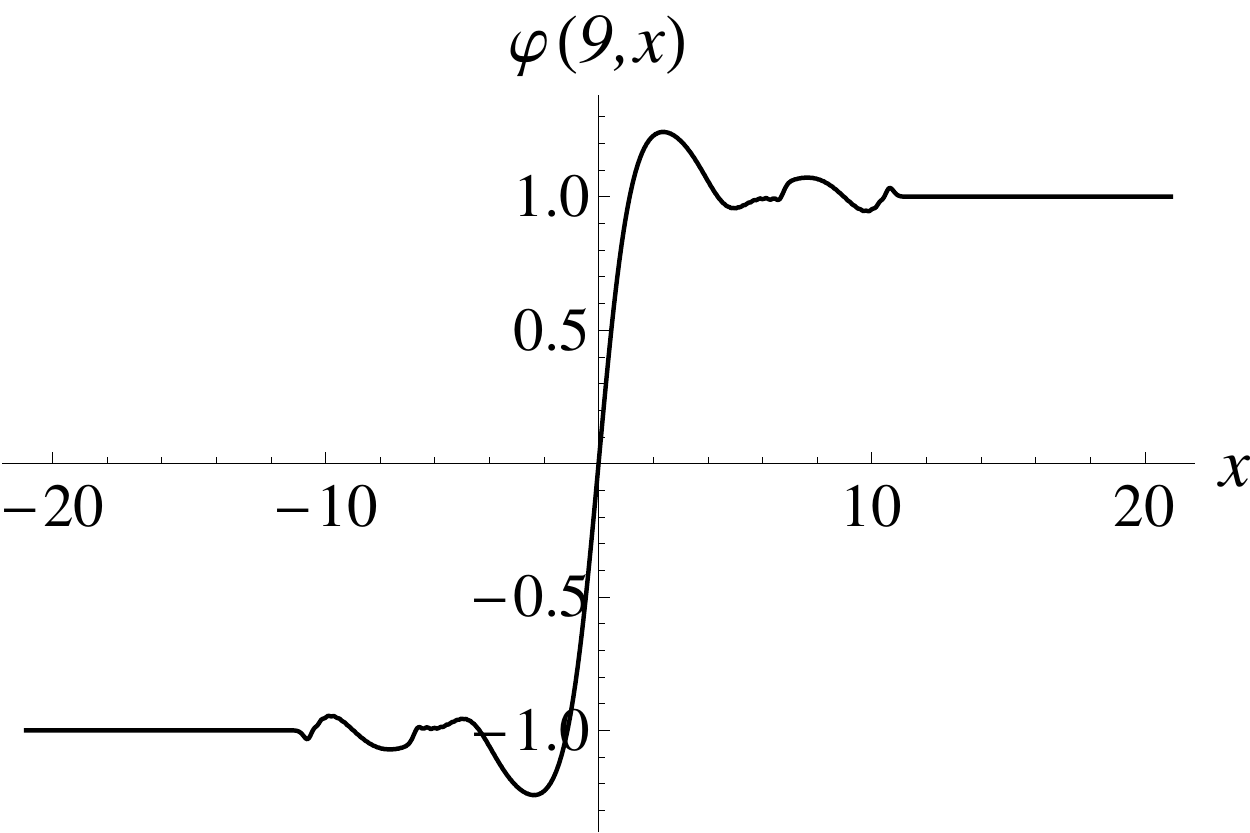}}
\end{minipage}
\hfill
\begin{minipage}[h]{0.49\linewidth}
\center{\includegraphics[width=0.83\linewidth]{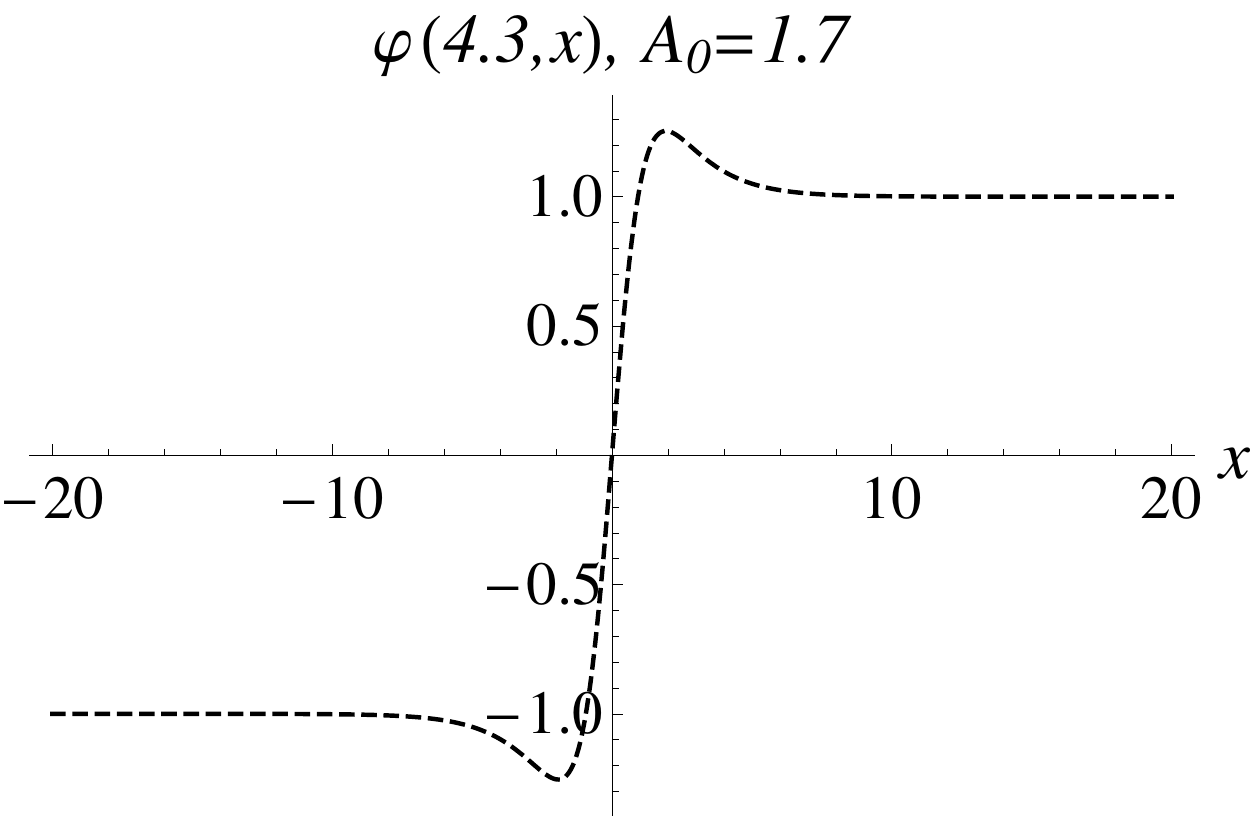}}
\end{minipage}
\caption{The profile of $\varphi(t,x)$ at $t=9$ (left panel) and mapped plot of the solution \eqref{eq:phiwithA} for $A_0=1.7$ at $t=4.3$ (right panel), $x_0=2$.}
\label{fig:sec3lev1pic5}
\end{figure}

In \cite{barashenkov}, it is shown that considering the substitution of $\varphi_{\scriptsize \mbox{K}}+\delta\varphi$ in Eq.~\eqref{eq:eom1} in the quadratic approximation by $\delta\varphi$ gives us an asymptotically stable solution. Its large amplitude vibrations are characterized by strong suppression. In Figs.~\ref{fig:sec3lev1pic3} and \ref{fig:sec3lev1pic5} two parts of the evolution and a comparison with the analytical solution \eqref{eq:phiwithA} are shown for two chosen moments of time.

For high values of $x_0$ the observed loops in the evolution are characterized by not-small amplitudes. In this case the final states of evolution can be identify with the elliptic solution $\varphi_{\scriptsize \mbox{el}}$ between two halves of the kink.


\subsection{\label{sec:sec3lev2} Result for an elliptic function with $\boldsymbol{0<\varphi_0<1}$}

\subsubsection{\label{sec:sec3lev2A} Dynamical initial state ($\delta\varphi_0<0$), configuration $(-1,\varphi_0,-1)$}

\begin{figure}[h]
\begin{minipage}[h]{0.49\linewidth}
\center{\includegraphics[width=0.83\linewidth]{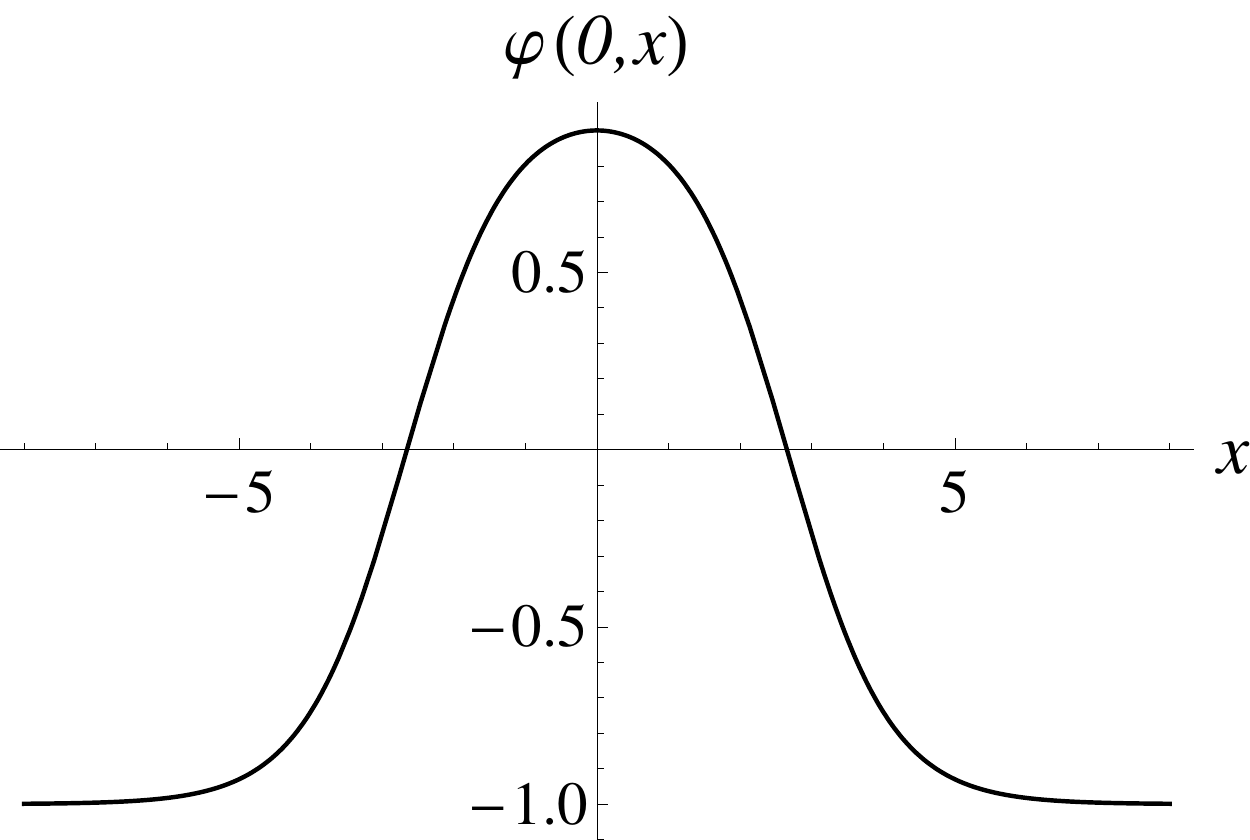}}
\end{minipage}
\hfill
\begin{minipage}[h]{0.49\linewidth}
\center{\includegraphics[width=0.83\linewidth]{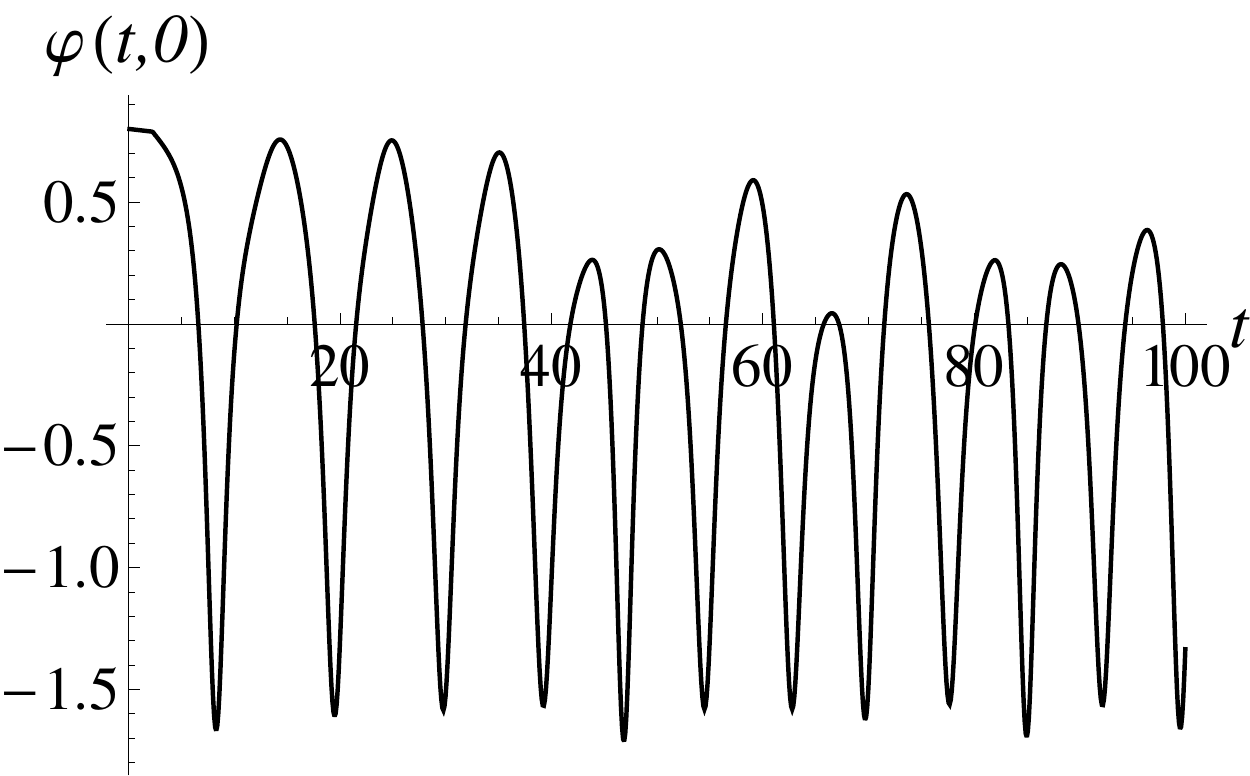}}
\end{minipage}
\caption{The profile of $\varphi(t,x)$ at $t=0$ (left panel) and the profile of $\varphi(t,0)$ (right panel), with parameters $\varphi_0=0.8$ and $\delta\varphi=-0.001$.}
\label{fig:sec3lev2pic1}
\end{figure}

In previous works a long-living configuration has been found, the so-called bion \cite{aek}. However, an analytical description of the observed process has not been given. In our work we take an initial condition $(-1,\varphi_0,-1)$, composed of one half-kink K and one half-antikink $\bar{\mbox{K}}$ as well as a half of period of $\varphi_{\scriptsize \mbox{el}}$ with fixed $\varphi_0$. This initial state is dynamical ($\varphi_0+\delta\varphi_0,\delta\varphi_0<0$) an is shown in Fig.~\ref{fig:sec3lev2pic1}. We obtain a long-living state with oscillation of the amplitude of $\varphi$ at $x=0$ (see Fig.~\ref{fig:sec3lev2pic1}). The observing oscillations in terms of an amplitude $\varphi_0(t)$ are called a regular bion. They also can be a new description of early found bion \cite{aek}.

\subsubsection{\label{sec:sec3lev2B} Statistic initial state ($\delta\varphi_0=0$), configuration $(-1,\varphi_0,0,-\varphi_0,1)$}

We take an initial condition $(-1,\varphi_0,0,-\varphi_0,1)$ for $\varphi_0>0.7$ (for smoother stitching), while the observed evolution does not qualitatively depend on $\varphi_0$. We also show the results for the case $\varphi_0=0.8$.

\begin{figure}[h]
\begin{minipage}[h]{0.49\linewidth}
\center{\includegraphics[width=0.83\linewidth]{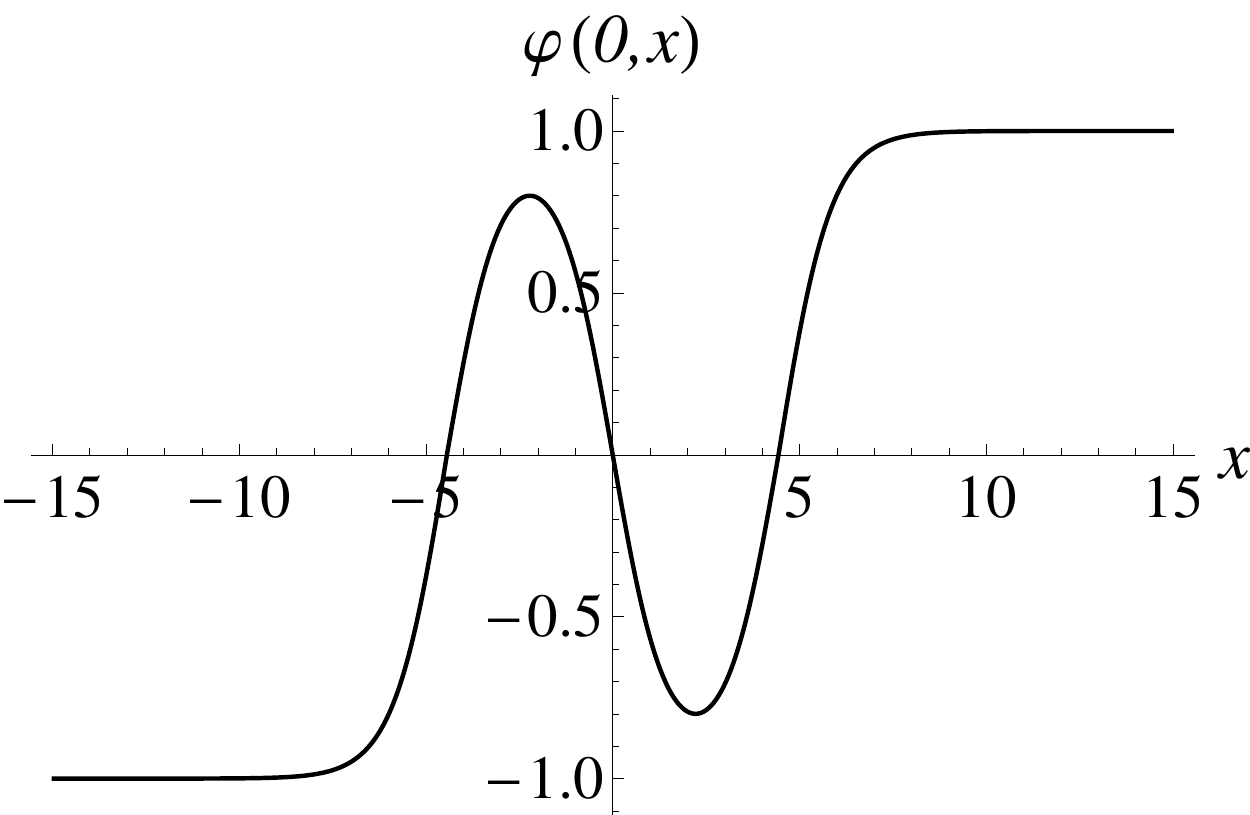}}
\end{minipage}
\hfill
\begin{minipage}[h]{0.49\linewidth}
\center{\includegraphics[width=0.83\linewidth]{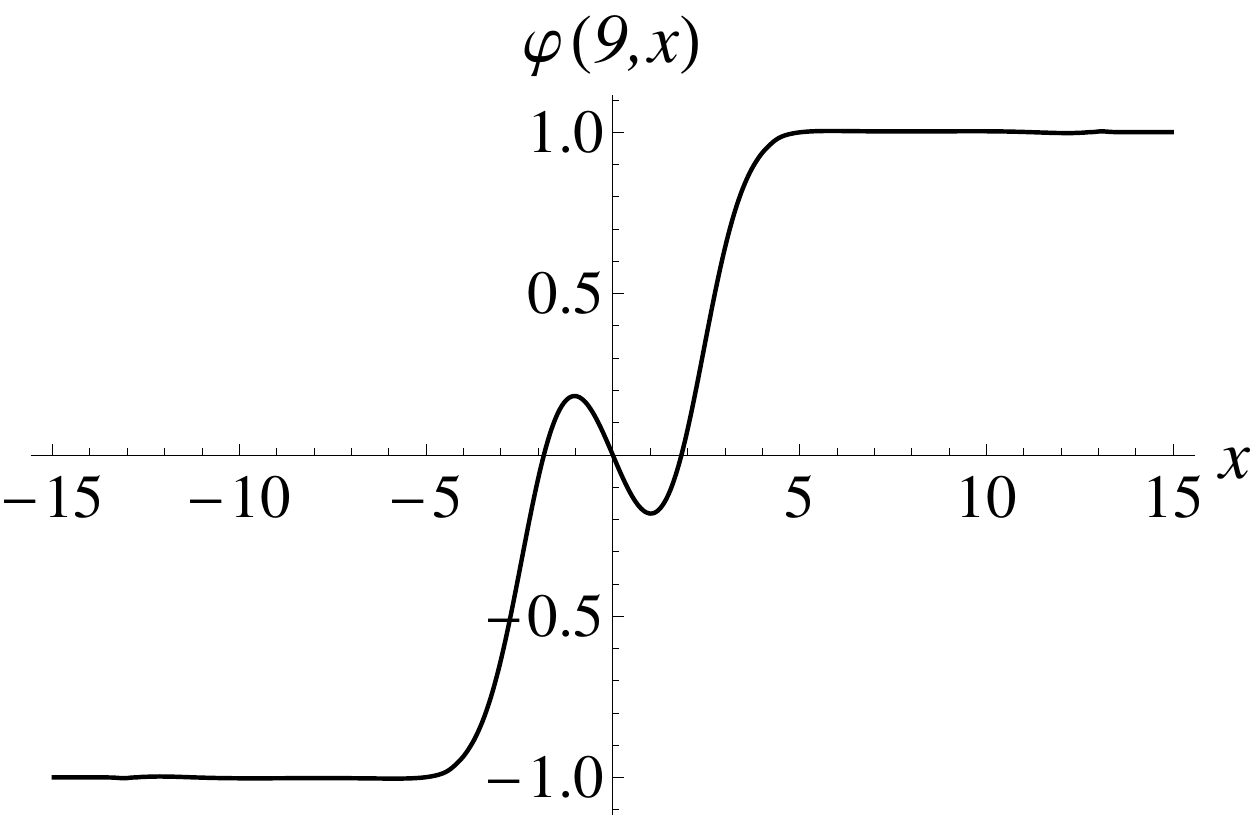}}
\end{minipage}
\caption{The profiles of $\varphi(t,x)$ at $t=0$ (left panel) and at $t=9$ (right panel). Parameters: $\varphi_0=0.8$ and $\delta\varphi_0=0$.}
\label{fig:sec3lev2Bpic1}
\end{figure}

\begin{figure}[h]
\begin{minipage}[h]{0.49\linewidth}
\center{\includegraphics[width=0.83\linewidth]{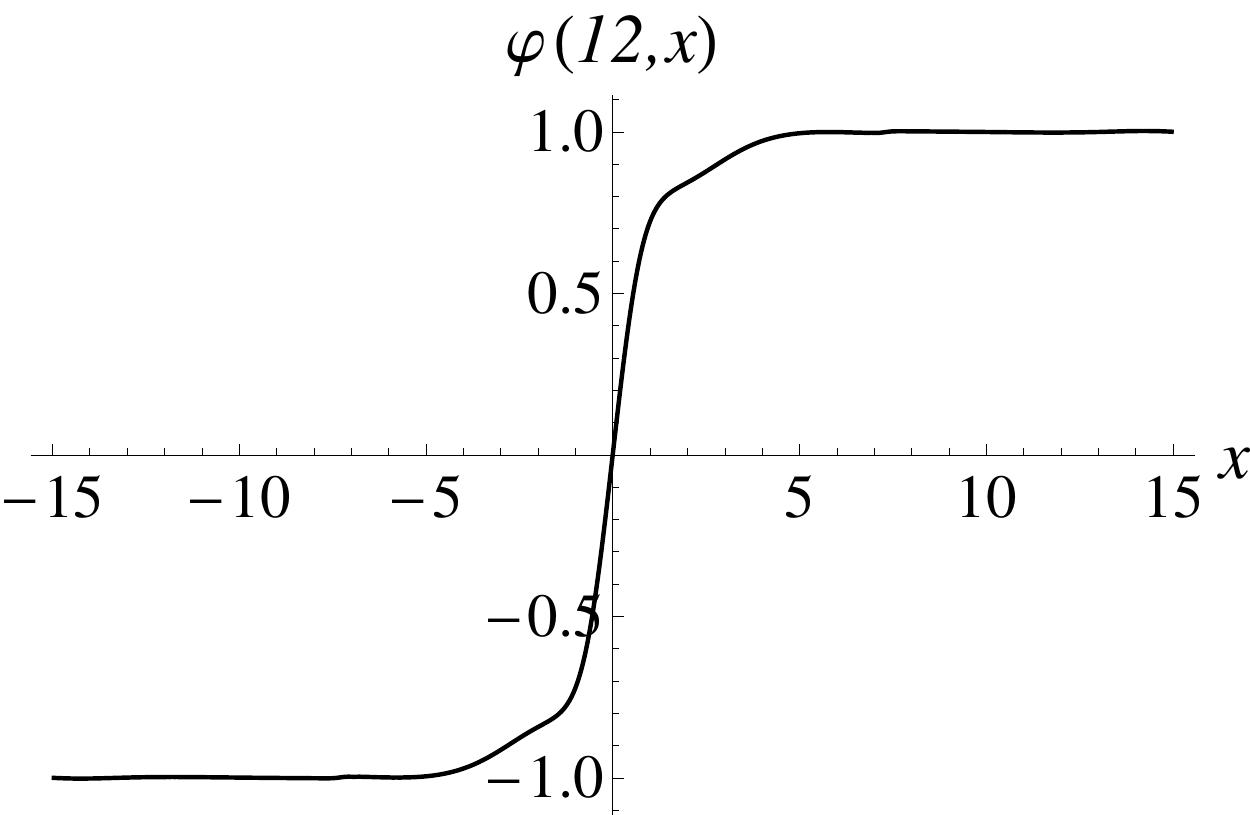}}
\end{minipage}
\hfill
\begin{minipage}[h]{0.49\linewidth}
\center{\includegraphics[width=0.83\linewidth]{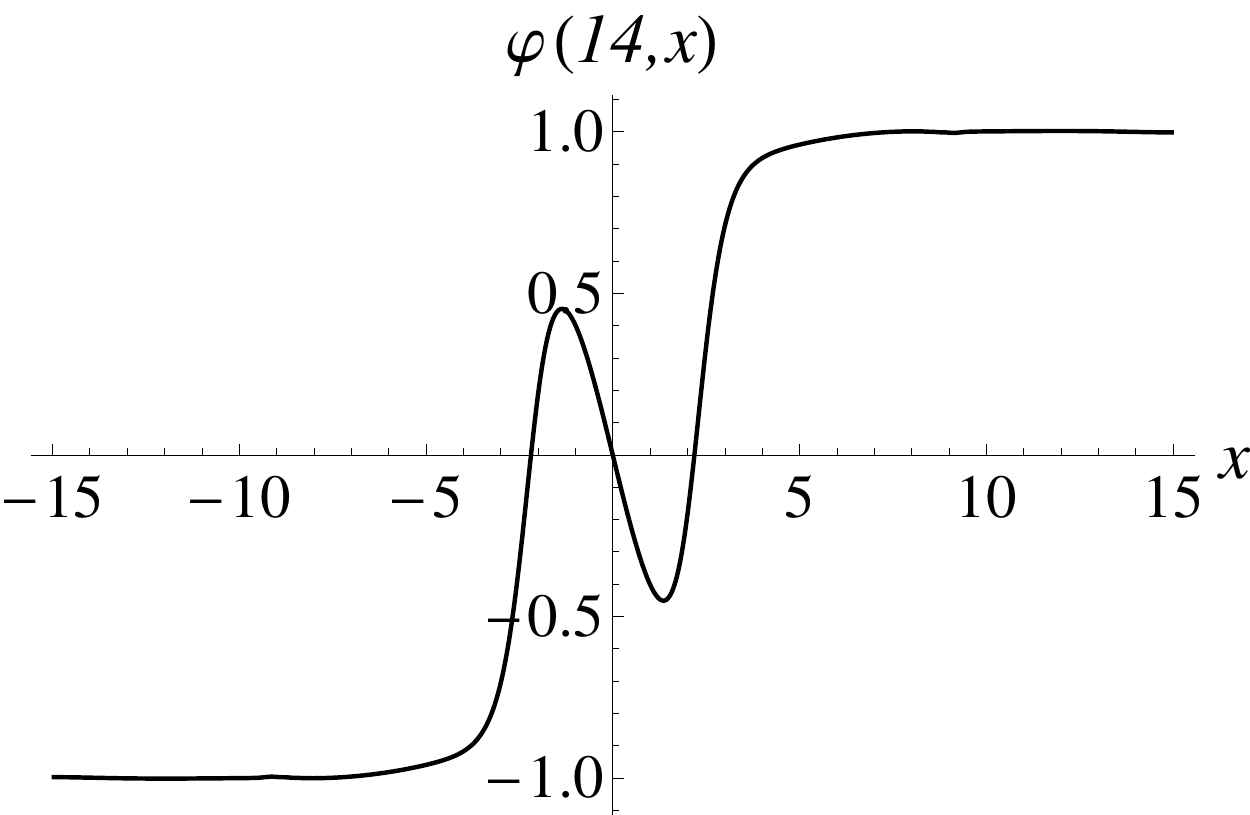}}
\end{minipage}
\caption{The profiles of $\varphi(t,x)$ at $t=12$ (left panel) and at $t=14$ (right panel). Parameters: $\varphi_0=0.8$ and $\delta\varphi_0=0$.}
\label{fig:sec3lev2Bpic3}
\end{figure}

We find two phases in the evolution: the external phase (a loop of high-amplitude formation) and the internal phase (a highly deformed kink). After some time the loops continue forming, but with smaller amplitude. After 4-5 cycles the external phase ends and the solution starts to resemble a long-living excited kink with the wave packet emission from the area of localization. This phenomenon is called a wobbling kink. This state is a final step of the evolution, which is observed for other variants of initial states. The profiles of $\varphi(t,x)$ for $\varphi_0=0.8$ at some particular time are shown in Figs.~\ref{fig:sec3lev2Bpic1} and \ref{fig:sec3lev2Bpic3}.


\subsubsection{\label{sec:sec3lev2D} Dynamical initial state ($\delta\varphi_0<0$), configuration $(-1,\varphi_0,0,-\varphi_0,1)$}

\begin{figure}[h!]
\begin{minipage}[h]{0.49\linewidth}
\center{\includegraphics[width=0.83\linewidth]{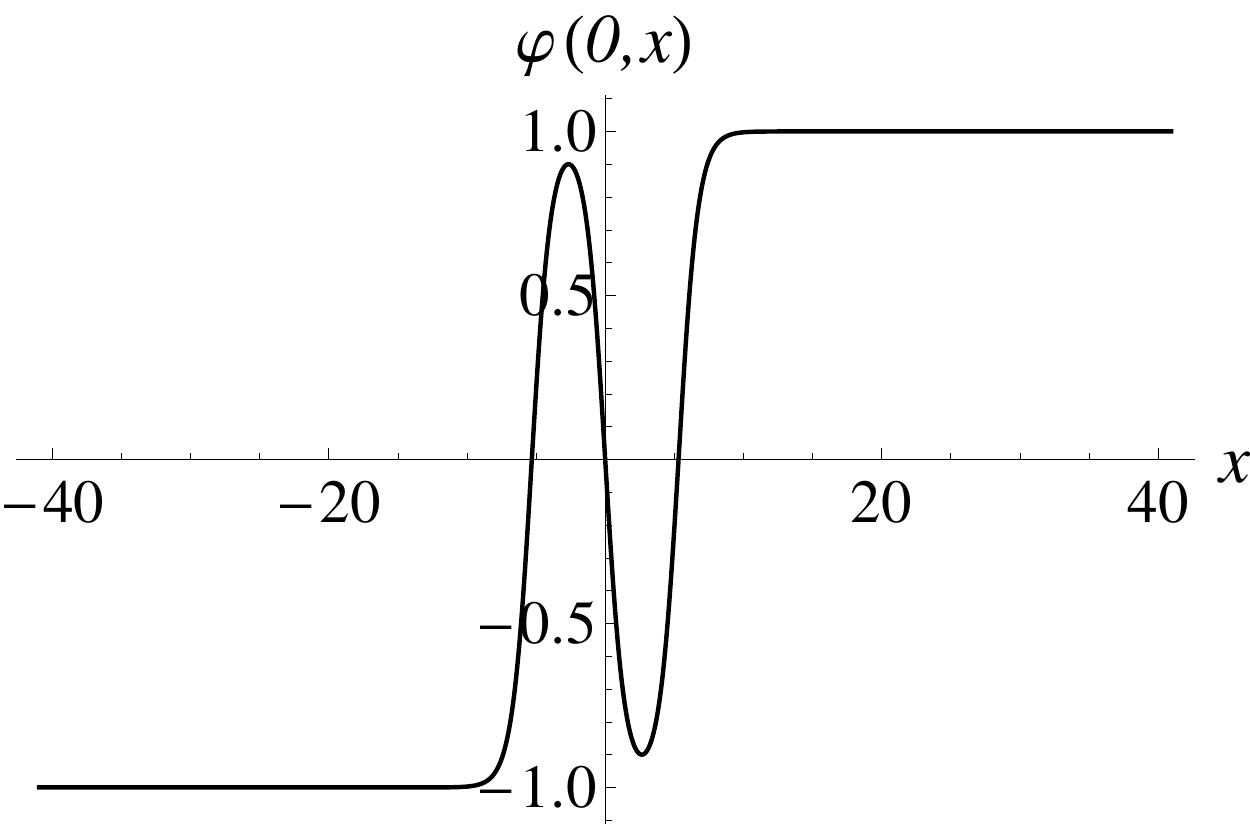}}
\end{minipage}
\hfill
\begin{minipage}[h]{0.49\linewidth}
\center{\includegraphics[width=0.83\linewidth]{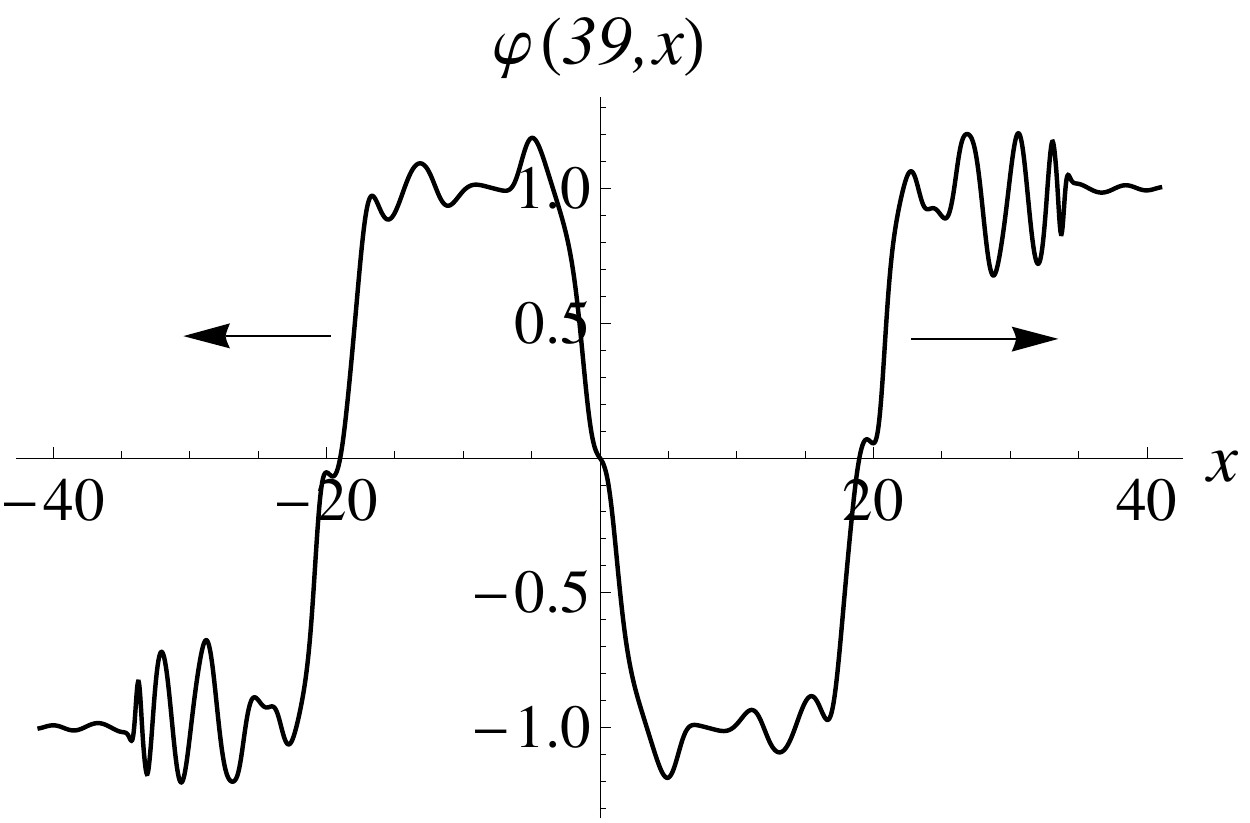}}
\end{minipage}
\caption{The profiles of $\varphi(t,x)$ at $t=0$ (left panel) and the formation of \eqref{eq:1step} at $t=39$ (right panel), with parameters $\varphi_0=0.9$ and $\delta\varphi_0=-0.0040$. The arrows indicate the direction of movement of formed kinks.}
\label{fig:sec3lev2Dpic4}
\end{figure}

\begin{figure}[h!]
\begin{minipage}[h]{0.49\linewidth}
\center{\includegraphics[width=0.83\linewidth]{sec3lev2Dpic7}}
\end{minipage}
\hfill
\begin{minipage}[h]{0.49\linewidth}
\center{\includegraphics[width=0.83\linewidth]{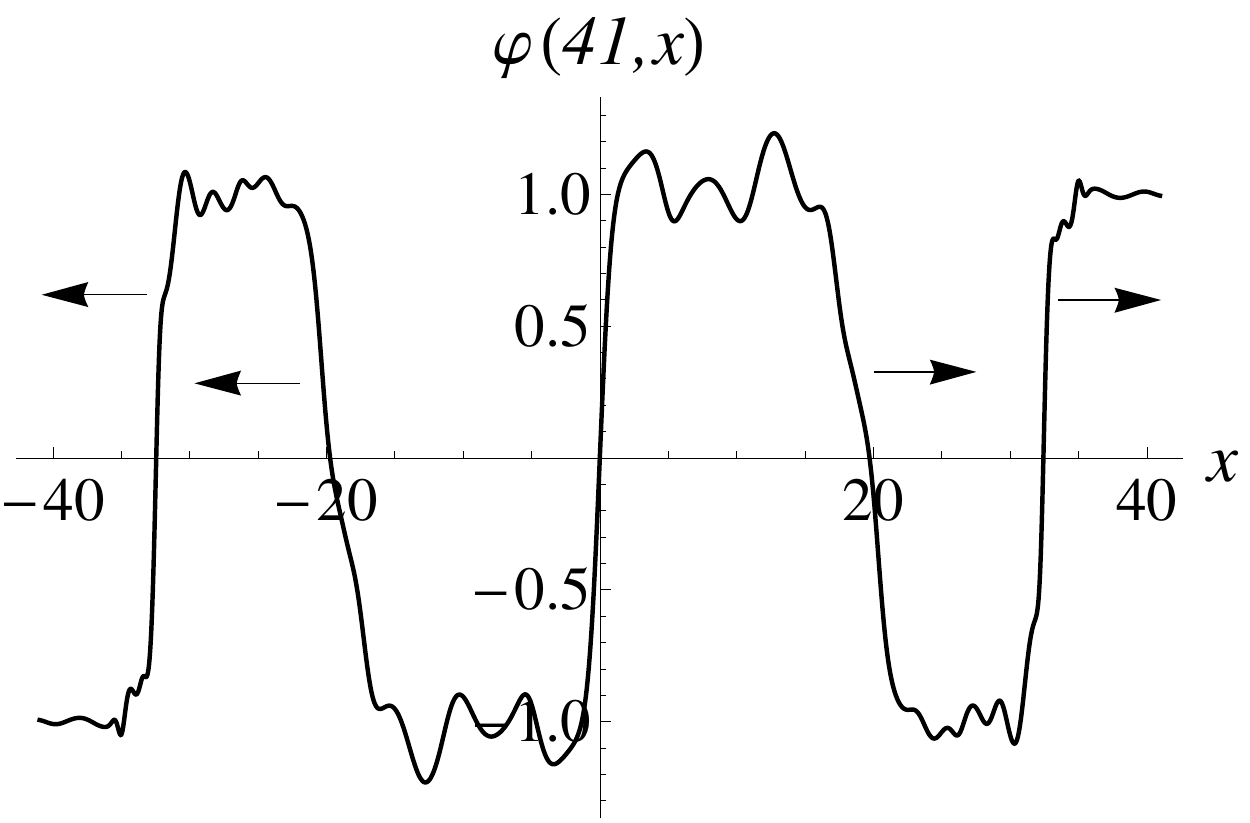}}
\end{minipage}
\caption{The profiles of $\varphi(t,x)$ at $t=0$ (left panel) and the formation of \eqref{eq:2step} at $t=41$ (right panel), with parameters $\varphi_0=0.9$ and $\delta\varphi_0=-0.0045$. The arrows indicate the direction of movement of formed (anti)kinks.}
\label{fig:sec3lev2Dpic5}
\end{figure}

An addition of $\delta\varphi_0<0$ to the initial state of the configuration $(-1,\varphi_0,0,-\varphi_0,1)$ leads to a faster reduction of the amplitude. At low values of $|\delta\varphi_0|,~ \delta\varphi_0<0$ the loops arise. For the first time during the evolution, the kink-antinkink pairs K$\overline{\mbox{K}}$ turn up. This phenomenon has a threshold. The increase of $|\delta\varphi_0|$ gives us a qualitatively new type of the evolution ($-0.0013<\delta\varphi_0<-0.0044$ for $\varphi_0=0.9$). In the system, we achieve
\begin{equation}\label{eq:1step}
\mbox{{\bf K}}\to\mbox{K}\overline{\mbox{{\bf K}}}\mbox{K},
\end{equation}
where in the center of the configuration a topological number is changing. The transition \eqref{eq:1step} is shown in Fig.~\ref{fig:sec3lev2Dpic4} for $\delta\varphi_0=-0.0040$. The next increasing of $|\delta\varphi_0|$ ($-0.0045\leq\delta\varphi_0<...$ for $\varphi_0=0.9$) gives us the next transition,
\begin{equation}\label{eq:2step}
\mbox{\bf K} \to\mbox{K}\overline{\mbox{K}}\mbox{{\bf K}}\overline{\mbox{K}}\mbox{K}.
\end{equation}
In this case we observe a conservation of topological number in the center. These transitions are observed for different $\varphi_0$. The transition \eqref{eq:2step} is shown in Fig.~\ref{fig:sec3lev2Dpic5} for $\delta\varphi_0=-0.0045$. We expect that with increasing the value of $|\delta\varphi_0|$, similar qualitative changes will be observed in the evolution of the initial state.


\section{Conclusions}

In this work we study new long-living solutions in the classical $\lambda\varphi^4$ field theory model in $(1+1)$ dimensions.

We use the cut-and-match method for forming initial states for numerical simulations. Using this method gives us new long-living solutions both for vacuum solutions and solutions with nontrivial topological number.

In previous work \cite{lit38}, a long-living configuration was observed in the kink-antikink scattering and was called a bion. In current work the cut-and-match method gives us an opportunity to describe a bion formation in a new qualitative way.

Furthermore, the highly excited states of the kink are observed in a sector with nontrivial topological number. We find a number of ways to reset this energy from this state. Except for emission of wave packet with small amplitude, firstly, an arising of the kink-antikink pairs has been observed. This phenomenon can perceived as a way to reset energy. At the same time there is a change of the topological number of the kink located in the central zone in the area. At lower excitation energies there is a long-living excited vibrational state of the kink. The phenomenon called the wobbling kink is final state for all considered initial conditions. After some time the excited state of a kink turns to a linearized one, which was formerly known as a discrete mode of exciting kink.

Despite the large number of new results, the cut-and-match method has a number of remaining issues in its application to the $\lambda\varphi^4$ model. In particular, a more detailed study of the dynamic of the initial conditions for the case of $\delta\varphi_0<0$ will be interesting, because in the last case there is the phenomenon of the birth of new kink-antikink pairs.

In the conclusion, we note that this research could be useful in different area of physics and, in particular, could be implemented in the description of the early stages of the evolution of the Universe.

\section*{ACKNOWLEDGMENTS}

We are very grateful to Professor Dr. I.L. Bogolubsky  for numerous critical comments during the reading of the manuscript. This work is supported by the Russian Federation Government under the Grant No.~945 from 18.11.2011. M.A. Lizunova also acknowledges the support from the Dynasty Foundation and Edward Lozansky. This work is part of the Delta Institute for Theoretical Physics (DITP) consortium, a program of the Netherlands Organisation for Scientific Research (NWO) that is funded by the Dutch Ministry of Education, Culture and Science (OCW).

\end{document}